\documentclass{article}
\usepackage{spconf,amsmath,graphicx}

\usepackage{multirow}
\usepackage{xcolor} 
\usepackage{wrapfig}
\title{The RWTH ASR system for TED-LIUM release 2: \\ improving hybrid HMM with SpecAugment}
\name{Wei Zhou$^{1,2}$, Wilfried Michel$^{1,2}$, Kazuki Irie$^{1,*}$, Markus Kitza$^{1,2}$, Ralf Schl\"uter$^{1,2}$, Hermann Ney$^{1,2}$
\thanks{*Work conducted while the author was at RWTH Aachen.
Now with the Swiss AI Lab, IDSIA, USI \& SUPSI, 6928 Manno-Lugano, Switzerland.}
}

\address{
$^1$Human Language Technology and Pattern Recognition, Computer Science Department,\\
  RWTH Aachen University, 52074 Aachen, Germany \\
$^2$AppTek GmbH, 52062 Aachen, Germany}

\begin{document}
%\ninept
%
\maketitle
\begin{abstract}
We present a complete training pipeline to build a state-of-the-art hybrid HMM-based ASR system on the 2nd release of the TED-LIUM corpus. Data augmentation using SpecAugment is successfully applied to improve performance on top of our best SAT model using i-vectors. By investigating the effect of different maskings, we achieve improvements from SpecAugment on hybrid HMM models without increasing model size and training time. A subsequent sMBR training is applied to fine-tune the final acoustic model, and both LSTM and Transformer language models are trained and evaluated. Our best system achieves a 5.6\% WER on the test set, which outperforms the previous state-of-the-art by 27\% relative.
\end{abstract}
\vspace{-2mm}
\section{Introduction \& Related Work}
\label{sec:intro}
\vspace{-1mm}
One of the most common neural network (NN) based acoustic modeling methods is the hybrid hidden Markov model (HMM) approach \cite{Bourlard1993hybridhmm}, which still gives state-of-the-art performance, as recently shown for benchmarks like Librispeech \cite{Luescher2019libsp} and Switchboard \cite{Kitza2019swb}. Bi-directional long short-term memory \cite{Hochreiter1997lstm} (BLSTM)-HMM are widely used for acoustic modeling in hybrid HMM systems. Based on the alignment generated from a Gaussian mixture model (GMM)-HMM baseline, cross-entropy (CE) training is usually applied to train the baseline NN models. Additionally, speaker adaptive training (SAT) using i-vectors and sequence discriminative training, such as the state-level minimum Bayes risk (sMBR) \cite{Gibson2006MBR} criterion, are often applied for further improvements.

Language models (LM) based on LSTM \cite{sundermeyer2012lstm} have been widely applied to automatic speech recognition (ASR). Large improvements are observed for both hybrid HMM systems and end-to-end systems \cite{Luescher2019libsp}. Transformer \cite{transformer} based LMs are reported to further improve over LSTM LMs \cite{Irie2019transformerLM}. For hybrid HMM systems, they are usually applied in lattice rescoring \cite{Irie2019transformerLM}, but also may be used in single-pass search \cite{Eugen2019lstmlm1pass}.

SpecAugment~\cite{specaugment}, as a simple feature augmentation method, has been successfully applied to end-to-end speech recognition systems. With increased model size and training time, end-to-end systems benefit strongly from SpecAugment \cite{Zeyer2019asru,Karita2019espnet}, and large improvements are also reported for end-to-end speech translation \cite{Bahar2019specaug}. However, its effect on hybrid HMM systems has not been thoroughly studied yet. 

In this work, we describe a complete training pipeline to build a state-of-the-art hybrid HMM-based ASR system on the 2nd release of the TED-LIUM corpus \cite{tedlium2} (TED-LIUM-v2). We apply SpecAugment in our training pipeline and obtain further improvement over our best SAT model using i-vectors. By investigating the effect of different maskings on hybrid models,  we achieve improvements from SpecAugment without increasing model size and training time. And no additional effort on learning rate scheduling is needed. Subsequent sequence discriminative training using the sMBR criterion is used to fine-tune the final acoustic model. For language modeling, both LSTM and Transformer based LMs are trained and evaluated. Our best system outperforms the previous state-of-the-art by a large margin.

\vspace{-2mm}
\section{Baseline Acoustic Model}
\label{sec:am}
\vspace{-1mm}
\subsection{Basic setups}
\vspace{-1mm}
All acoustic models are trained on the 207 hours of training data of TED-LIUM-v2. The official dictionary with roughly 152k words and 160k pronunciations is used. To evaluate all intermediate acoustic models with recognition experiments, we use a fixed heavily pruned 4-gram LM, whose details are described in Sec.~\ref{sec:lm} denoted as `4-gram-small'. This allows us to simplify the tuning, and report the relative improvement of each step. All trainings are done using our NN modeling toolkit RETURNN \cite{returnn}, our ASR toolkit RASR \cite{rasr} and our work-flow manager Sisyphus \cite{sisyphus}. All recognition results are obtained with maximum a posteriori (MAP) Viterbi decoding.

\vspace{-2mm}
\subsection{Baseline}
\label{subsec:baseline}
\vspace{-1mm}
We follow the standard steps described in \cite{Luescher2019libsp} to train the GMM-HMM baseline. Starting with linear alignment, monophone GMMs are trained with 16-dimensional MFCCs and their first oder derivatives. With each triphone modeled by 3 HMM states, generalized triphone states are obtained by state tying using classification and regression tree (CART). We use 9k CART labels. Generalized triphone state GMMs are then trained on windowed MFCC with linear discriminant analysis (LDA) transformation. This step is repeated once to refine the CART labels with better alignment. Subsequently, vocal tract length normalization (VTLN) and SAT using constrained maximum likelihood linear regression are applied to further improve the GMMs. The final alignment from the VTLN-SAT GMM is used in the next step to train NN baseline with the CE criterion.

We use 80-dimensional logmel features for the NN training. The NN model contains six BLSTM layers with 512 units for each direction. This topology is used in all further steps. The Nesterov-accelerated adaptive moment estimation (Nadam) optimizer \cite{Nadam} with an initial learning rate of 0.0009 is used. Greedy layer-wise pre-training \cite{layerwisepretrain} and Newbob learning rate scheduling \cite{Zeyer17newbob} with a decay factor of 0.9 are applied. CE and focal loss \cite{Lin2017focalloss} with factor of 2 are used. The training set is split into 5 subepochs and models converge well with roughly 32 full epochs. Sequences are decomposed into chunks of 64 frames with 50\% overlap and a mini-batch of 128 chunks is used. Additionally, a 10\% dropout \cite{dropout} and $L_2$ regularization with a factor of $0.01$ are applied to all hidden layers. 

Table~\ref{tab:basewer} shows the word error rate (WER) results of each of the aforementioned training steps. We also try to use the BLSTM baseline to generate a new alignment and repeat the NN training, but no further improvement is obtained. This new alignment is used in further steps of training.

\begin{table}
\caption{\it WERs(\%) of baseline acoustic models (evaluated with the 4-gram-small LM on the dev set)}
\begin{center}
\label{tab:basewer}
\begin{tabular}{|c|c|l|c|}
\hline
Unit & Model           & Feature      & Dev \\ \hline
monophone  & \multirow{5}{*}{GMM}   & MFCC         & 41.6 \\ \cline{1-1}\cline{3-4}
\multirow{6}{*}{triphone} &  & + LDA        & 21.8 \\ \cline{3-4}
                       &                           & \quad + VTLN & 21.0 \\ \cline{3-4}
                       &                           & \quad + SAT & 19.5 \\ \cline{3-4}
                       &                           & \quad\quad + VTLN & 19.4 \\ \cline{2-4}
 & \multirow{2}{*}{BLSTM}                          & logmel       & 10.4 \\ \cline{3-4}
                       &                           & \quad + i-vectors & \phantom{1}9.8 \\ \hline
\end{tabular}
\end{center}
\vspace{-7mm}
\end{table}

\vspace{-2mm}
\subsection{I-Vector Adaptation}
\label{subsec:ivec}
\vspace{-1mm}
We follow \cite{Kitza2019swb} to apply SAT using i-vectors as speaker embeddings. The embeddings are concatenated to the logmel features at each frame. The universal background model (UBM) is trained on the whole training set. To train the UBM, logmel features with a context of 9 frames are concatenated and then reduced to a dimension of 60 with LDA. I-vectors are then estimated for each recording separately using all feature frames including non-speech. We follow \cite{Kitza2019swb} to use a size of 100 for the i-vectors. As shown in the last two rows of Table~\ref{tab:basewer}, 6\% relative improvement is achieved by applying SAT with i-vectors. We expect to achieve larger improvements with further tuning of the embedding parameters.

\vspace{-2mm}
\section{SpecAugment}
\vspace{-1mm}
The original SpecAugment \cite{specaugment} applies time warping, time masking and frequency masking on logmel features. Since time warping is reported to have minor effect, we skip it in our training. This also avoids the additional effort to handle the alignment accordingly. We apply the two maskings on logmel features concatenated with i-vectors. Since i-vectors are included, the frequency masking is renamed as feature masking. Both maskings are bounded to the fixed chunk size and feature dimension, and are realized in a similar way as described in~\cite{Bahar2019specaug}.

\vspace{-2mm}
\subsection{Time Masking (TM)}
\vspace{-1mm}
With a chunk of $T$ frames $(x_1,...,x_T)$, a position $t$ is randomly selected from $[1,T]$. Then a time mask of length $\Delta t$ is randomly selected from $[0,\Delta t_{\text{max}}]$, where $\Delta t_{\text{max}}$ is a predefined maximum time mask length. TM is then applied by setting the consecutive frames $(x_t,...,x_{t+\Delta t})$ to $0$. This procedure is repeated $m$ times, where $m$ is randomly selected from $[1,M]$. $M$ is a predefined maximum iteration number for TM. Thus, TM can be controlled by setting $M$ and $\Delta t_{\text{max}}$ accordingly, which we denote as $M \times \Delta t_{\text{max}}$.

\vspace{-2mm}
\subsection{Feature Masking (FM)}
\vspace{-1mm}
With features of dimension $D$, an index $d$ is randomly selected from $[1,D]$. Then a feature mask of length $\Delta d$ is randomly selected from $[0,\Delta d_{\text{max}}]$, where $\Delta d_{\text{max}}$ is a predefined maximum feature mask length. FM is then applied by setting the features within dimension $[d,d+\Delta d]$ to $0$. This procedure is again repeated $n$ times, where $n$ is randomly selected from $[1,N]$. $N$ is a predefined maximum iteration number for FM. Similar to TM, FM can be controlled by setting $N$ and $\Delta d_{\text{max}}$ accordingly, which we denote as $N \times \Delta d_{\text{max}}$.

\vspace{-3mm}
\subsection{SpecAugment on Logmel with I-Vectors}
\vspace{-1mm}
To further improve the previous best baseline model, we directly apply the TM and FM on the 80-dimensional logmel features concatenated with 100-dimensional i-vectors. The random selections in both TM and FM are independently applied for each chunk in a batch. BLSTM models are trained from scratch. The predefined $M$ and $N$ are halved in the first 2000 steps for a more stable pre-training. We set a default $\Delta d_{\text{max}}$ to 10\% of the feature dimension. With $D=180$, this means $\Delta d_{\text{max}}=18$. Then a default $N=5$ is used for a maximum of 50\% FM (denoted as $5\times18$). For hybrid HMM systems, CART labels consume much less frames than label units used in end-to-end systems. By setting a very large $\Delta t_{\text{max}}$, evidence of several continuous CART labels are masked out, which might be less beneficial. Therefore, we start with a default $\Delta t_{\text{max}}=5$ to match the maximum duration of a speech CART label based on our experience. Due to the fixed chunk size $T=64$, $M$ also has to be limited to keep a reasonable ratio of TM. With $\Delta t_{\text{max}}=5$, we set a default $M=6$ for roughly a maximum of 50\% TM (denoted as $6\times5$). 

We first investigate the effect of different TM with the default FM ($5\times18$). Under the same maximum ratio of TM, we compare a set of different $M \times \Delta t_{\text{max}}: \{15\times2, 6\times5, 3\times10, 2\times15, 1\times30\}$ to find the optimal $\Delta t_{\text{max}}$. As shown in Table~\ref{tab:specaugmentwer}, too long TM gives less improvement, which matches our expectation. Surprisingly too short TM is also less beneficial, which should be resulted from the decreased effect of TM. With the optimal $\Delta t_{\text{max}}=10$, we further optimize $M$ by training models with $M: \{2, 4\}$ to apply less and more TM (denoted as $2\times10$ and $4\times10$). From Table~\ref{tab:specaugmentwer} we see neither of them brings further improvement. The best result achieves 7\% relative improvement over the SAT baseline using i-vectors.

We then investigate the effect of different FM with the best TM ($3\times10$). Similarly under the same 50\% maximum ratio of FM, we compare a set of different $N \times \Delta d_{\text{max}}: \{10\times9, 5\times18, 3\times30\}$ to find the optimal $\Delta d_{\text{max}}$. As shown in Table~\ref{tab:specaugmentwer}, our default FM setting still gives the best result. With optimal $\Delta d_{\text{max}}=18$, we train models with $N:\{3, 7\}$ to vary the maximum ratio of FM (denoted as $3\times18$ and $7\times18$). Both give worse results. Additionally, we also investigate the importance of i-vectors in terms of FM. Since they are fixed for each frame, we train a model with default FM applied only within the logmel features and i-vectors are left untouched. This is reflected by the column 'FM on Ivec' in Table~\ref{tab:specaugmentwer}. For 80-dimensional logmel features, 10\% of the feature dimension results in $\Delta d_{\text{max}}=8$. The result of applying FM only within logmel features is much worse. This shows that including i-vectors for FM is essential, which brings certain variation also into the speaker features.

Finally, we also investigate the effect to continue training the i-vectors-based SAT baseline with SpecAugment. We use the best masking settings obtained so far, i.e. TM=$3\times10$ and FM=$5\times18$. In this case, we turn off the pre-training and its corresponding 2000 steps of halved masking. The learning rate is reset to allow an escape from local optimum. The model converges slightly faster than training from scratch directly with SpecAugment, but it only achieves the same performance of 9.1\% WER in the end. Considering the much longer training time in total, there is not too much benefit to follow this track.

\begin{table}
\caption{\it WERs(\%) of further training steps based on logmel features concatenated with i-vectors (evaluated with the 4-gram-small LM on the dev set)}
\vspace{-2mm}
\begin{center}
\setlength{\tabcolsep}{0.4em}
\label{tab:specaugmentwer}
\begin{tabular}{|c|c|c|c|c|r|}
\hline
\multirow{2}{*}{Criterion} & \multicolumn{3}{c|}{SpecAugment}  & \multirow{2}{*}{Dev} \\ \cline{2-4}
                           & FM on Ivec & $N \times \Delta d_{\text{max}}$ & $M\times\Delta t_{\text{max}}$ &  \\ \hline \hline
\multirow{12}{*}{\hspace{-4mm}CE}        & \multicolumn{3}{c|}{none} & 9.8 \\ \cline{2-5}
  & \multirow{10}{*}{yes} & \multirow{7}{*}{\hspace{3mm}$5\times18$} & $15\times2$ & 9.5 \\ \cline{4-5}
  & & & \hspace{2mm}$6\times5$ & 9.3 \\ \cline{4-5}
  & & & \hspace{4mm}$3\times10$ & \textbf{9.1} \\ \cline{4-5}
  & & & \hspace{4mm}$2\times15$ & 9.2 \\ \cline{4-5}
  & & & \hspace{4mm}$1\times30$ & 9.4 \\ \cline{4-5}
  & & & \hspace{4mm}$2\times10$ & 9.3 \\ \cline{4-5}  
  & & & \hspace{4mm}$4\times10$ & 9.2 \\ \cline{3-5}
  & & $10\times9$ & \multirow{5}{*}{\hspace{4mm}$3\times10$} & 9.5 \\ \cline{3-3} \cline{5-5}
  & & \hspace{4mm}$3\times30$ & & 9.2 \\ \cline{3-3} \cline{5-5} 
  & & \hspace{4mm}$3\times18$ & & 9.3 \\ \cline{3-3} \cline{5-5}
  & & \hspace{4mm}$7\times18$ & & 9.4 \\ \cline{2-3} \cline{5-5}
  & no & \hspace{2mm}$5\times8$ & & 9.6 \\ \hline
\quad + sMBR & \multicolumn{3}{c|}{none} & \textbf{8.6} \\ \hline
\end{tabular}
\end{center}
\vspace{-6mm}
\end{table}

\vspace{-3mm}
\subsection{Discussion}
\vspace{-2mm}
Overall, the improvements from SpecAugment are not as large as reported for end-to-end systems \cite{specaugment}. However, improvements are obtained without increasing model size and training time. Models converge well with roughly the same number of epochs as needed for the baseline training. Additionally, no careful design of learning rate scheduling is needed (only Newbob is applied here), although more improvements can be explored by doing this.

In general, end-to-end systems need larger amounts of training data to be competitive with state-of-the-art hybrid HMM systems. This situation is eased by training end-to-end systems with SpecAugment for many more epochs. Together with the results in this work, we tend to infer that in terms of SpecAugment, end-to-end systems benefit most from having more data, whereas hybrid HMM benefit from more variation introduced into the data. However, more investigation is needed for a thorough understanding.

\vspace{-2mm}
\begin{table}[!h]
\vspace{-1mm}
	\setlength{\tabcolsep}{0.3em}  % horizontal space inside cells.
	\centering
	\caption{\it Perplexity of the word-level LMs.
		The same 152K vocabulary is used for all models (except for the small 4-gram which contains 52 words less).}
	\label{tab:ppl}
	\begin{tabular}{ |l|r|r|r|} \hline
		\multirow{2}{*}{Model}  & Param & \multicolumn{2}{|c|}{PPL}    \\ \cline{3-4}
		& in M. & \multicolumn{1}{|c|}{Dev} & \multicolumn{1}{|c|}{Test}    \\ \hline
		%4-gram     &  343 & 105.4 & 124.7 \\ \hline
  4-gram-small &  4   &  135.0   & 169.9  \\ \hline
		4-gram & 161 & 113.2 & 127.9 \\ \hline
		LSTM   & 450 & 73.5& 71.3 \\ \hline
		Transformer  &  414  &  \textbf{62.0} & \textbf{60.7} \\  \hline
	\end{tabular}
	\vspace{-2mm}
\end{table}

\begin{table*}
\centering
\caption{\it WERs(\%) of the final acoustic model with different language models on both dev and test sets of TED-LIUM-v2, and a summary of most relevant results from the literature}
\label{tab:smbrwer}
\begin{tabular}{|c|c|c|c|c|r|r|}
\hline
\multirow{2}{*}{Paper} & \multicolumn{2}{c|}{Acoustic Model} & \multicolumn{2}{c|}{Language Model} & \multirow{2}{*}{Dev} & \multirow{2}{*}{Test} \\ \cline{2-5}
                                  & Approach & \multicolumn{2}{c|}{Labels} & Approach & & \\ \hline \hline
Zeyer et al. \cite{Zeyer2019asru} & \multirow{2}{*}{E2E} & \multicolumn{2}{c|}{BPE} & Transformer & 10.3 & 8.8 \\ \cline{1-1} \cline{3-7}
Karita et al. \cite{Karita2019espnet} &  & \multicolumn{2}{c|}{SentencePiece} & RNN & 9.3  & 8.1 \\ \hline
Han et al. \cite{Han2019multistride}  & \multirow{6}{*}{hybrid HMM} & \multirow{6}{*}{triphone} & \multirow{6}{*}{word} & 4-gram & 7.7 & 8.0 \\ \cline{1-1} \cline{5-7}
\multirow{2}{*}{Han et al. \cite{Han2018capio}} &  &  &  & 4-gram & 7.6  & 8.1 \\ \cline{5-7}
                                                &  &  &  & RNN & 7.1  & 7.7 \\ \cline{1-1} \cline{5-7}
\multirow{3}{*}{this work}            &                             &  &  & 4-gram & 6.8 & 7.3 \\ \cline{5-7}
                                      &                             &  &  & LSTM & \textbf{5.6} & \textbf{6.0} \\ \cline{5-7}
                                      &                             &  &  & Transformer & \textbf{5.1} & \textbf{5.6} \\ \hline
\end{tabular}
\vspace{-2mm}
\end{table*}

\vspace{-2mm}
\section{Sequence Discriminative Training}
\vspace{-2mm}
We follow \cite{Luescher2019libsp} to further apply sequence discriminative training on the best model from the previous step. In this case, we take the SAT model using i-vectors trained with the best SpecAugment setting. We use a lattice-based version of sMBR training criterion to fine-tune the model weights. No SpecAugment is applied in this step. This converged CE model and a bi-gram LM trained on the TED-LIUM-v2 LM training data are used for lattice generation and initialization of model training. We then continue training with a small constant learning rate of $1\times10^{-5}$ and use early stopping to prevent overfitting on the training data. CE smoothing with a scale of 0.1 is applied. As shown in Table \ref{tab:specaugmentwer}, the sequence discriminative training achieves an additional 6\% relative improvement.

\vspace{-1mm}
\section{Language Modeling}
\label{sec:lm}
\vspace{-2mm}
The LM training data consists of 7 subsets including the TED-LIUM-v2 training audio transcriptions, with a total of 270\,M running words. The small 4-gram LM is trained in a similar way as the Kaldi example recipe \cite{kaldi}. All the rest of our LMs have been described in \cite{irie20}. We refer readers interested in more details to this paper. 
 
We first train modified Kneser-Ney 4-gram language models \cite{kneser1995improved, chen1999empirical, sundermeyer:lmsmoothing:is2011} on each subset of the training data with the word level vocabulary of size 152K. We linearly interpolate these sub-LMs including a background 4-gram model trained on all training text, using the interpolation weights optimized for the development perplexity.

We train both LSTM and Transformer language models. The LSTM LM has 4 layers with 2048 nodes in each layer. The Transformer model has 32 layers with a feed-forward inner dimension of 4096, a self-attention embedding dimension of 768, and 12 attention heads per layer.
No positional encoding is used.
The input word embedding dimension is 128 for both models. Table \ref{tab:ppl} shows the corresponding perplexities.

\vspace{-1mm}
\section{Experimental Results}
\vspace{-2mm}
The final acoustic model trained with the sMBR criterion is evaluated with better language models. LM scales are optimized on the development set. A one-pass recognition setup with MAP Viterbi decoding is applied for both the 4-gram LM and the LSTM LM, where the generated lattices from the LSTM LM-based recognition are used for lattice rescoring with the Transformer LM.

Tabel~\ref{tab:smbrwer} shows the WER results of these experiments together with a brief summary of best results from the literature. These include hybrid HMM systems as well as end-to-end (E2E) systems using different model types, topologies and label units, such as byte pair encoding (BPE) and SentencePiece \cite{Kudo2018sentencepiece}. We refer readers to the original papers for more details. As shown in the table, the previous best system \cite{Han2018capio} has a 7.7\% WER on the test set. Our best result is 5.6\% on the test set, which achieves 27\% relative improvement.

\vspace{-4mm}
\section{Conclusion}
\vspace{-2mm}
In this work, we presented the integration of data augmentation using SpecAugment into the training pipeline of a state-of-the-art ASR system based on hybrid HMM approach for the TED-LIUM-v2 corpus. SpecAugment provides 7\% relative improvement on top of our best SAT model using i-vectors, more precisely from 9.8\% to 9.1\% WER on the dev set with a small 4-gram LM. We analyzed the effect of different maskings and found out that SpecAugment is beneficial in all cases. The major impact comes from the maximum time and feature mask lengths, which have to be optimized. Then with a good control of maximum ratio of TM and FM, decent improvements are achieved without increasing model size and training time. For feature masking, it is essential to include all features even if i-vectors are fixed for each frame of the segment. Additionally, we found that training from scratch with SpecAugment directly is more efficient than continuing training with SpecAugment to achieve similar performance. Together with subsequent sMBR training and Transformer LM, our best hybrid HMM system achieves the state-of-the-art performance with 5.6\% WER on the test set, which improves over the previous best WER of 7.7\% by 27\% relative.

\vspace{-4mm}
\section{Acknowledgements}
\vspace{-3mm}
\begin{wrapfigure}[5]{l}{0.13\textwidth}
	\vspace{-4mm}
	\begin{center}
		\includegraphics[width=0.15\textwidth]{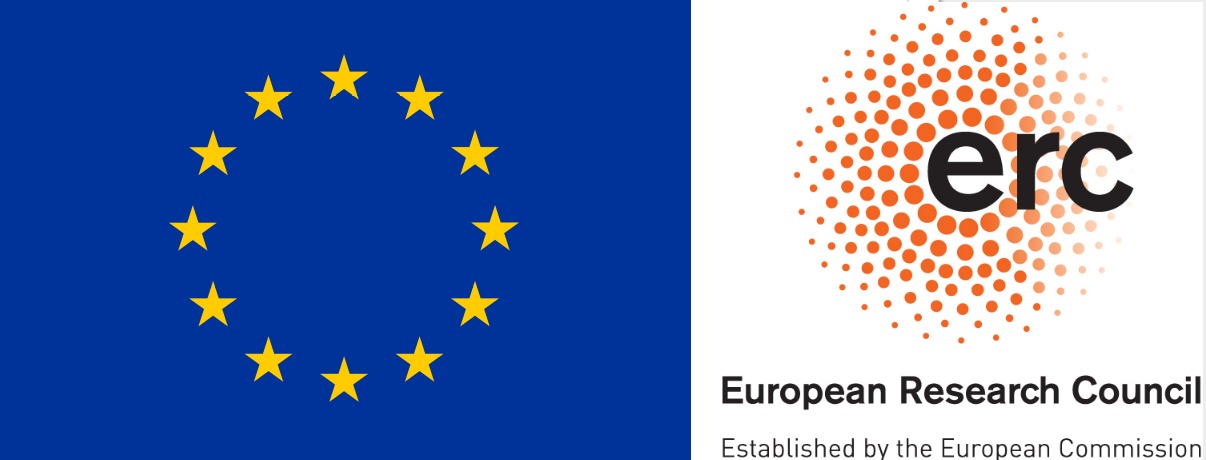} \\
	\end{center}
	\vspace{-4mm}
\end{wrapfigure}
\small
This work has received funding from the European Research Council (ERC) under the European Union's Horizon 2020 research and innovation program (grant agreement No 694537, project ``SEQCLAS") and from a Google Focused Award. The work reflects only the authors' views and none of the funding parties is responsible for any use that may be made of the information it contains.

We thank Albert Zeyer, Christoph L\"{u}scher, Pavel Golik, Peter Vieting and Tobias Menne for useful discussions.
%\vspace{-2mm}

\let\normalsize\small\normalsize
% http://tex.stackexchange.com/questions/93859/condense-the-space-between-bibliographic-entries
\let\OLDthebibliography\thebibliography
\renewcommand\thebibliography[1]{
	\OLDthebibliography{#1}
	\setlength{\parskip}{-0.3pt}
	\setlength{\itemsep}{1pt plus 0.07ex}
}

\bibliographystyle{IEEEbib}
\bibliography{refs}

\end{document}